# Detailed Composition of Stars in Dwarf Spheroidal Galaxies

MATTHEW D. SHETRONE
*University of Texas, McDonald Observatory*

## 1.1 Introduction

Simulations of galaxy formation in a hierarchical scenario suggest that the halo of the Milky Way (MW) may have formed from many smaller protogalaxies and that the protogalaxy building blocks farthest from the center of the MW gravitational potential may have turned into small galaxies such as the Local Group dwarf spheroidal (dSph) system. The age and abundance estimates of the dSph systems suggest that these dSph systems have metallicities and ages similar to that of the MW halo. High-resolution abundance analysis of dSph giants is a fairly new field of study that grew out of the advent of 8–10 m class telescopes. The dSph systems are generally old, and thus the brightest stars available for high-resolution study are the giants near the tip of the giant branch. This type of study is similar to the early studies of globular clusters. The similarity with globular clusters cannot be overemphasized. Many of the advances in chemical evolution come from using the globular clusters as templates of how chemical evolution occurs. However, nearly all globular clusters are mono-metallic and thus only exhibit the abundance patterns one might expect for a specific metallicity (and a specific star formation rate). From early studies of the most nearby dSph galaxies, it has been known that they are not mono-metallic and thus could serve as even more important test cases for our knowledge of chemical and stellar evolution.

In this review the research done to date on high-resolution dSph abundances is summarized, including several not-yet-refereed studies, and some connection to our knowledge of chemical evolution and hierarchical galaxy formation is drawn. The current tally of surveyed galaxies is given in Table 1.1. As can be seen from Table 1.1 different researchers take different tactics in their approach to investigating dSph abundances. The Shetrone et al. teams' spectra have just a few spectra in each dSph galaxy with moderate signal-to-noise ratio (S/N) spectra, while the Smecker-Hane and Bonifacio teams have concentrated on a large sample in the closest dSph with high S/N in each spectrum. Both approaches have merit and yield different insights into the chemical evolution of the dSph system. For this review the dSph galaxies have been divided into two categories—simple and complex—based upon their implied star formation histories (e.g., Dolphin 2002). Draco, Ursa Minor, Sextans, and Sculptor we classify as simple because the implied star formation history has a single burst followed by a decline in star formation. Leo I, Fornax, Carina, and Sagittarius (Sgr) we classify as complex because their implied star formation histories contain more than a single burst of star formation. Other than the overall metallicity typified by Fe, we will concentrate on five classes of elements: the light $\alpha$ elements (represented by O and Mg), the heavy $\alpha$ elements (represented by Ca and Ti), the light odd-Z elements (represented by Na), a few iron-peak elements (represented by Mn and Cu), and the neutron capture elements (represented by Y, La, Ba, and Eu).

Table 1.1. *The High-resolution dSph Sample*

| dSph | # Stars | Observatory | Reference |
|---|---|---|---|
| Sculptor | 5 | VLT UVES | Shetrone et al. (2003) |
|  | 4 | VLT UVES | Geisler et al. (2004) |
| Ursa Minor | 3 | Keck HIRES | Shetrone, Bolte, & Stetson (1998) |
|  | 3 | Keck HIRES | Shetrone, Côté, & Sargent (2001) |
| Draco | 4 | Keck HIRES | Shetrone, Bolte, & Stetson (1998) |
|  | 2 | Keck HIRES | Shetrone, Côté, & Sargent (2001) |
|  | 1 | Keck HIRES | Fulbright, Rich, & Castro (2004) |
| Sextans | 5 | Keck HIRES | Shetrone, Côté, & Sargent (2001) |
| Sagittarius | 2 | VLT UVES | Bonifacio et al. (2000) |
|  | 14 | Keck HIRES | Smecker-Hane & McWilliam (2004) |
|  | 10 | VLT UVES | Bonifacio & Caffau (2003) |
|  |  |  | Bonifacio et al. (2003) |
| Carina | 5 | VLT UVES | Shetrone et al. (2003) |
| Fornax | 3 | VLT UVES | Shetrone et al. (2003) |
| Leo I | 2 | VLT UVES | Shetrone et al. (2003) |

## 1.2 Dangers and Caveats

One of the main dangers in constructing a coherent data set from different analyses is the zeropoints created by the observing and analysis techniques of different researchers. Because the dSph stars are observed with 8 m-class telescopes and telescope time is quite precious, there is almost no overlap between samples. However, there is a single exception: in the sample of Shetrone et al. (1998) there was a Draco star observed with very low metallicity, D119. This star was later reobserved by Fulbright et al. (2004). The Fulbright spectrum of D119 had much higher S/N ($\sim$100) than that of Shetrone et al. ($\sim$40). The techniques employed to get effective temperature, metallicity and surface gravity are different for these two authors. Despite all of these differences the abundance ratios are in reasonable agreement: within 1 $\sigma$ for Fe, Ca, Ti, and Cr, and 2 $\sigma$ for Na, Mg, and Ni. What is important about the results from the higher-S/N data in Fulbright et al. are the smaller error bars and interesting upper limits that can be used in chemical evolution analysis.

One of the easiest, although tedious, corrections that has to be made when combining the analysis of different authors is the choice of oscillator strengths. This is easy when the same lines are used by different authors but nearly impossible when different line sets are used. In this review some attempt will be made to bring the line lists into a common system. Examples of where this will have a strong impact include the Mg I $\lambda$5528 Å abundances (Shetrone et al. 2001 corrected upward by 0.14 dex and Geisler et al. 2004 and Bonifacio et al. 2000 corrected downward by 0.1 dex), the Mg I $\lambda$5711 Å abundances (Geisler et al. 2004 corrected downward by 0.11 dex), the Ca I $\lambda$6156 Å abundances (Bonifacio et al. 2000 corrected up-ward by 0.29 dex), the Ca I $\lambda$6161 Å abundances (Bonifacio et al. 2000 corrected upward by 0.25 dex), the Ca I $\lambda$6166 Å abundances (Bonifacio et al. 2000 corrected upward by 0.24 dex), and the Ca I $\lambda$6508 Å abundances (Bonifacio et al. 2000 corrected upward by 0.39 dex). Where there is little overlap in the choice of lines, the person combining the data sets must be wary and look for abundance jumps or inconsistencies in the abundance trends.

Hyperfine splitting (HFS) is also a correction that not all analyses include. For example, Shetrone et al. (1998), Bonifacio et al. (2000), and Shetrone et al. (2001) did not include HFS for Ba, Eu, Cu or Mn, while Smecker-Hane & McWilliam (2004) and Shetrone et al. (2003) did. HFS affects different elements differently and even different lines within the same species are affected differently; for example, Mn HFS is quite large and should be included for all lines, while the HFS for the red Ba lines are not very large, and the blue lines of Ba have larger HFS corrections. Generally speaking strong lines are affected more than weak lines, and thus metal-poor stars have smaller HFS corrections than metal-rich stars. For this review, corrections for HFS will be made when access to the data, published or unpublished, is available.

Differences in analysis techniques and model atmospheres used by the different researchers can also lead to zeropoint differences in the different data sets. In the work of Shetrone et al. (2003) some attempt was made to quantify the abundance differences using the ATLAS/WIDTH vs. MARCS/MOOG; they found that the abundances changed by typically 0.06 dex, with no change being larger than 0.12 dex. Shetrone et al. (1998, 2001, 2003) have employed MARCS/MOOG, Smecker-Hane & McWilliam (2004) have employed ATLAS/MOOG, and Bonifacio et al. (2000, 2004) and Bonifacio & Caffau (2003) have employed ATLAS/WIDTH. No attempt will be made in this review to compensate for differences in analysis techniques.

Non-local thermodynamic equilibrium (NLTE) modeling is not a standard part of the abundance analysis of dSph stars at the present time. This is in part due to the comparative nature of the analysis. The researchers compare the abundances of the dSph stars to similar analysis of MW stars, and since these MW stars were not analyzed with NLTE, neither are the dSph stars. No attempt to make NLTE corrections has been made in this review.

## 1.3  Simple Systems

The best studied among the simple systems is Sculptor (Scl), with five stars from Shetrone et al. (2003) and four stars from Geisler et al. (2004), and as such we begin our review by looking in detail at this galaxy as a template for the other simple systems. Figure 1.1 exhibits the heavy $\alpha$ elements (Mg and O) abundance ratios plotted against metallicity. The solid line represents a MW toy model where Fe yields of Type II supernovae (SNe II) and Type Ia supernovae (SNe Ia) are included and a constant star formation rate is assumed. The yields in this toy model are set by matching the MW abundances and do not rely on the theoretical SN yields. When the SNe Ia begin to contribute, the slope changes but the $\alpha$ abundance continues to rise because the star formation rate has not changed and SNe II continue to contribute $\alpha$ elements. The most metal-poor Scl dSph stars fall on the MW toy model, while the more metal-rich stars fall below that model. The dashed line represents a Scl toy model (we do not intend to imply that this model is correct and only use it for discussion purposes). The Scl toy model is similar to the MW toy model except that at an arbitrary point ([Fe/H] = −1.8, which is approximately the peak metallicity for many of the simple dSph systems) the star formation rate is dropped by a factor of 10 (another arbitrary value set to fit Fig. 1.1 only). In all other respects the Scl toy model will use the same parameters as the MW toy model (i.e. yields set to fit the MW abundance pattern). The rapid decline in [O/Fe] and [Mg/Fe] after [Fe/H] = $-1.8$ is due to the onset of SNe Ia, and since SNe Ia create very little light $\alpha$ elements and large amounts of Fe, these ratios decline. This can also be seen in the right panels of Figure 1.1, where the $\alpha$ abundances are plotted against metallicity. The total amount of O and Mg in the most metal-rich stars is nearly the same as the stars that have nearly 6 times lower [Fe/H]. Figure 1.2 shows a similar plot for the heavy $\alpha$ elements. In this figure the Ca and Ti abundances are not constant over the [Fe/H] range $-1.8$ to $-1.0$: there is an increase in the total amount of Ca and Ti. This is possible if SNe Ia produce a small amount of the heavy $\alpha$ elements (e.g., Iwamoto et al. 1999).

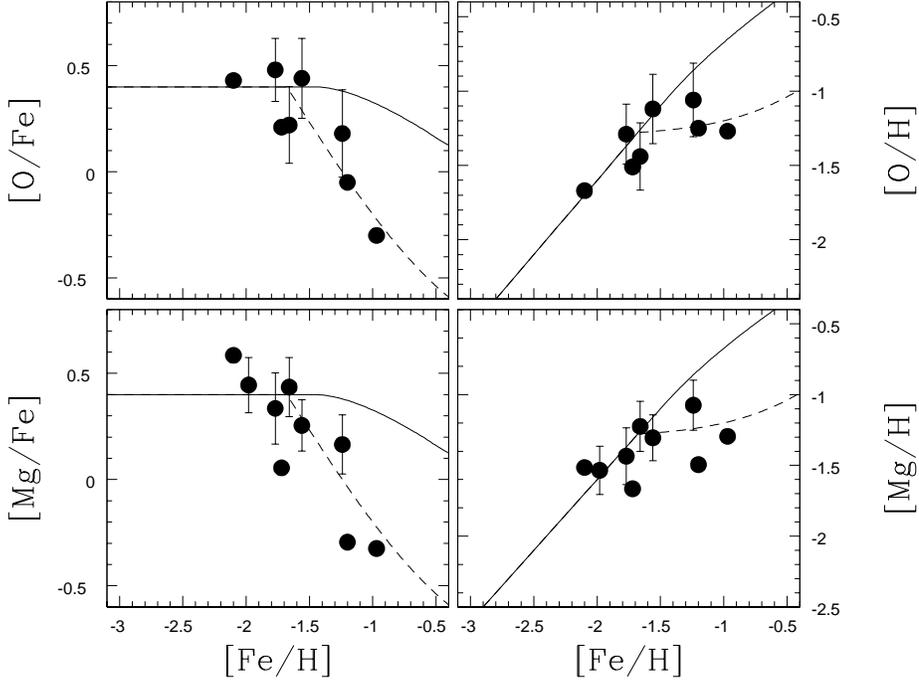

Fig. 1.1. The filled circles with error bars are from Shetrone et al. (2003), and the ones without error bars are from Geisler et al. (2004). The solid line represents the MW toy model and MW mean. The dashed line represents a Scl toy model.

In Figure 1.3 the Eu abundances and *s*-process to *r*-process ratio ([Ba/Eu]) are plotted against the metallicity. Again the solid line represents the MW toy model, and the dashed line represents the Scl toy model. In both models the *s*-process time scale is taken to be the same as the SN Ia time scale (the Eu yields in this toy model do not include any metallicity terms), and no contribution from the *s*-process is included in Eu (Eu is nearly a pure *r*-process element even at [Ba/Eu] = 0). In the MW toy model the Eu abundance continues to increase as the SNe Ia begin to contribute Fe because the star formation rate is held constant and the SNe II continue to produce Eu. In the middle panel the ratio of *s*-process to *r*-process (Ba to Eu) are shown, and the models show a clear transition from a nearly pure *r*-process ratio at low metallicity to a mixture of *r*-process and *s*-process at higher metallicity. Recall that at [Fe/H] = −1.8 the star formation rate in the Scl toy model drops by a factor of 10, and thus the Eu abundance (largely produced by SNe II) should not increase. The most metal-rich star has a very large Eu abundance and is potentially self-contaminated or contaminated from a binary companion. In the Scl toy model, [Ba/Eu] rapidly rises, as the asymptotic giant branch (AGB) stars begin to contribute *s*-process material, and fits the Scl data reasonably well. The lowest panel of Figure 1.3 suggests that the Eu and light $\alpha$ elements are produced on the same time scales, as is also seen in the MW.

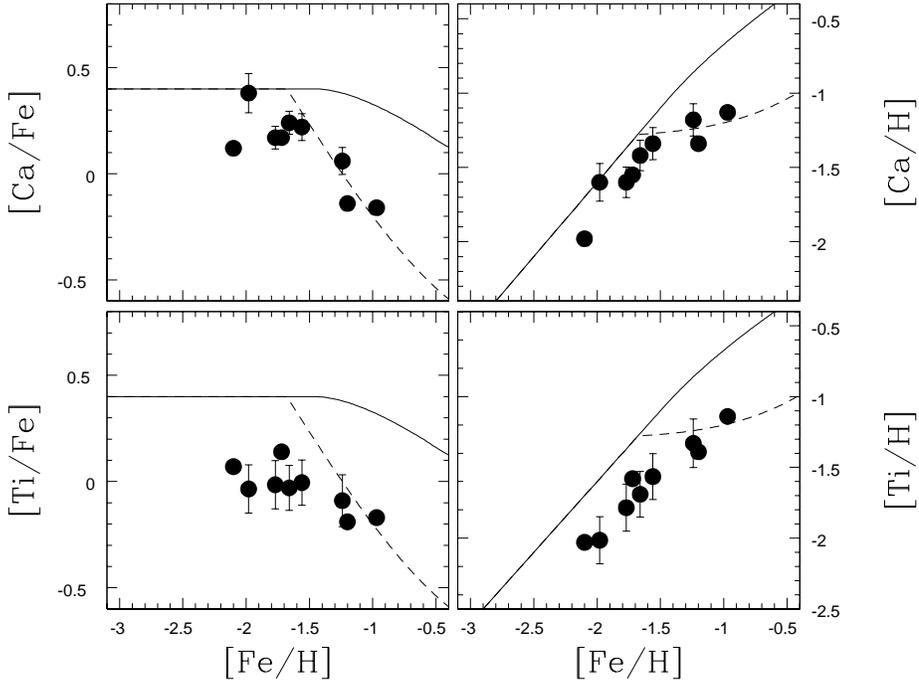

Fig. 1.2. The symbols are the same as in Fig. 1.1.

The understanding of the origin of Y is poorly understood in the MW. In very metal-poor stars observations suggest that there may be an additional source of Y (and Sr) in addition to the *r*-process that produced the heavier *r*-process elements (e.g., Ryan, Norris, & Beers 1996; Burris et al. 2000). This has led to models that split the production of light *r*-process elements and heavy *r*-process elements (e.g., Qian & Wasserburg 2001). Because of the complex nature of the Y abundances we have not tried to create a simple model and instead plot the MW abundances as a reference in Figure 1.4. For the more metal-poor stars the Y abundances behave in a way similar to the MW halo, although the [Ba/Y] abundances may be high with respect to the halo. For the more metal-rich Scl stars, where the *s*-process should begin to contribute, the results are a bit mixed. Two of the Y abundances are systematically lower than the MW halo, and [Ba/Y] is correspondingly larger, while one is similar to the MW. With only three stars it is difficult to determine what is the actual trend. In Figure 1.4 all of the simple dSph data are shown, and the trend of higher [Ba/Y] at higher metallicities becomes clear. The average [Ba/Y] for the simple dSph stars with [Fe/H] $< -1.8$ is 0.46 dex, while the average [Ba/Y] for the simple dSph stars with [Fe/H] $> -1.8$ is marginally larger at 0.62 dex. As was pointed out by Smecker-Hane & McWilliam (2004), these high values for heavy *s*-process to light *s*-process are typical for metal-poor AGB stars (e.g., Busso et al. 2001). What remains a mystery is why the pure *r*-process [Ba/Y] in the dSph is 0.46 dex, while the MW halo stars have an average [Ba/Y] of −0.08 dex, based on the Fulbright (2002) data.

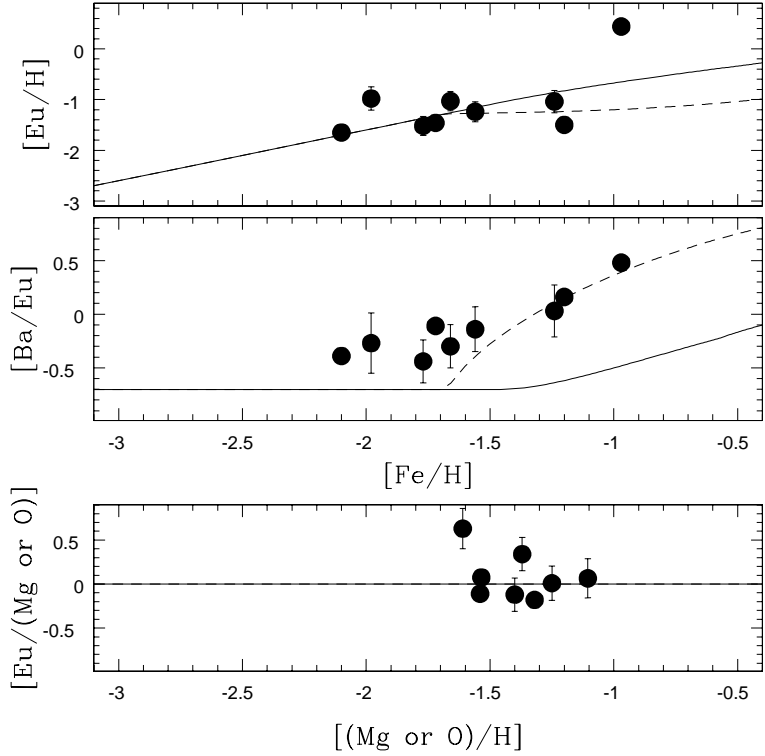

Fig. 1.3. The symbols are the same as in Fig. 1.1.

In our Scl toy model we predict that there is significant contribution of SNe Ia in the most metal-rich stars. This leads to the prediction that the Mn and Cu abundances for these stars should be significantly enhanced *if* SNe Ia are responsible for the upturn seen at [Fe/H] in the MW. Figure 1.5 exhibits [Mn/Fe] and [Cu/Fe] plotted against the metallicity. The solid line represents the MW toy model, the dotted line represents a pure SN II contribution based on the metallicity-dependent yields of Woosley & Weaver (1995), and the dashed line represents the Scl toy model. The Scl toy model uses the same prescription as the MW toy model, whereby Cu and Mn are produced in SNe Ia. Clearly the Scl toy model is a poor fit to the data, while the metallicity-dependent SN II model is somewhat better. A similar argument can be made for any of the other simple dSph Mn and Cu abundances, although care should be given to the Shetrone et al. (2001) data since no correction for HFS was made that would bring the Mn and Cu abundances *down* preferentially for the strongest lines (i.e. the most metal-rich stars). These data eliminate the possibilities of metal-poor SNe Ia as a significant contributor to the rise in [Mn/Fe] seen in the MW. Further constraints on the nucleosynthesis of Mn are introduced by examining the Mn abundances in the more metal-rich Sgr dSph and MW bulge stars (McWilliam, Rich, & Smecker-Hane 2003). Using these different environments, they conclude that Mn yields from both SNe Ia and SNe II are metallicity dependent. Detailed modeling will be required to eliminate one of these two possibilities.

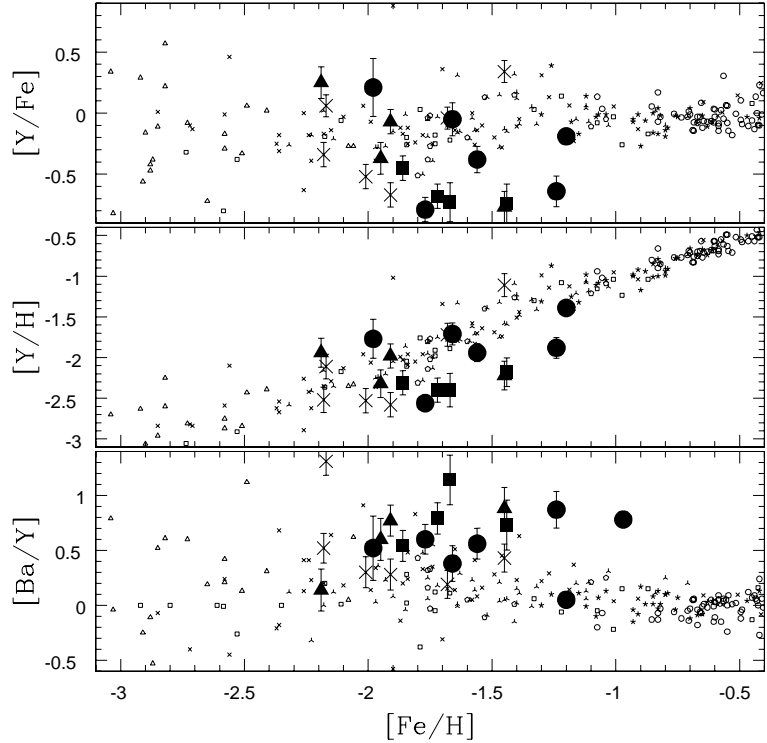

Fig. 1.4. The filled circles are the same as in Fig. 1.1. The large symbols with error bars are from Shetrone et al. (2001): the filled squares are Draco stars, filled triangles are Sextans stars, and the crosses are Ursa Minor stars. The small symbols without error bars are from the literature MW field halo sample: Gratton & Sneden (1988, 1991, 1994), Edvardsson et al. (1993), McWilliam et al. (1995), Nissen & Schuster (1997), Stephens (1999), Burris et al. (2000), Prochaska et al. (2000), and Stephens & Boesgaard (2002).

The Scl Na abundance ratios generally fall at the low end of the MW abundance distribution. Smecker-Hane & McWilliam (2004) have attributed the low Na abundances in the Sgr dSph galaxies to metallicity-dependent SN II yields and an excess of Fe from SNe Ia that should not produce any Na (e.g., Thielemann, Nomoto, & Yokoi 1986), the latter being dominant in Sgr. This explanation works well for the more metal-rich dSph stars, but leads to the question of why the more metal-poor stars in the simple dSphs do not have halo-like Na abundances. Taking the entire simple dSph abundance set between [Fe/H] = −3 to −1.9, we find an average [Na/Fe] = −0.30 ($\sigma = 0.24$), while the Fulbright (2002) halo stars in the same metallicity range have an average [Na/Fe] = −0.05 ($\sigma = 0.27$). The dSph stars fall near the bottom of the MW [Na/Fe] distribution, although a slight trend with metallicity cannot be ruled out due to the low number of stars near [Fe/H] = −3.

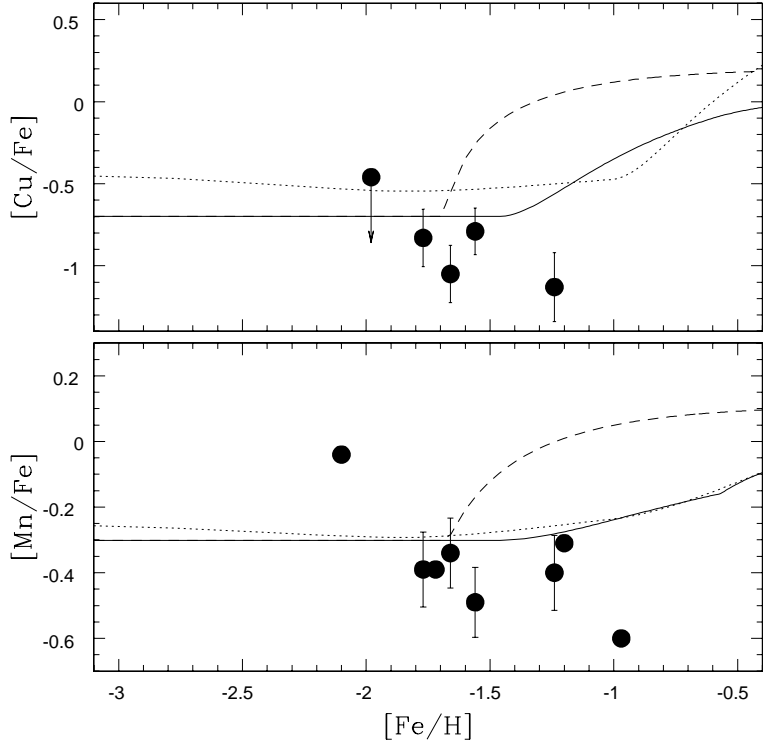

Fig. 1.5. The symbols are the same as in Fig. 1.1. The dotted line is a pure SN II toy model with yields from Woosley & Weaver (1995).

The toy model fails to explain one important feature of the abundances in Scl: the heavy $\alpha$ elements among the most metal-poor stars in Scl fall systematically below the prediction. This can be seen in the left panes of Figure 1.2, where the [Ca/Fe] and [Ti/Fe] abundance ratios are plotted against the metallicity. The SN II model predictions are based on MW halo abundances. Only a single star in the Scl sample has a Ca abundance that falls in the mid range of what is found in the MW halo, and that star has the largest error bar in the sample. All of the other Scl abundances fall at the bottom of the MW Ca distribution at all metallicities. This sub-halo abundance pattern can be seen in the other simple dSph systems as well (see Fig. 1.6). In previous investigations the $\alpha$ elements were grouped together to form an "$\alpha$" abundance, and the difference between the light $\alpha$ elements and the heavy $\alpha$ elements was not stated explicitly. Fulbright et al. (2004) suggested the split between the light and heavy $\alpha$ elements is mimicked by the most massive SN II yields, as seen in the Woosley & Weaver (1995) yields. Thus, preferentially keeping high-mass SN II yields by changing the initial mass function, invoking selective mass loss from the galaxy, or creating a different prompt inventory for dSph galaxies could create the observed split between the light and heavy $\alpha$ elements.

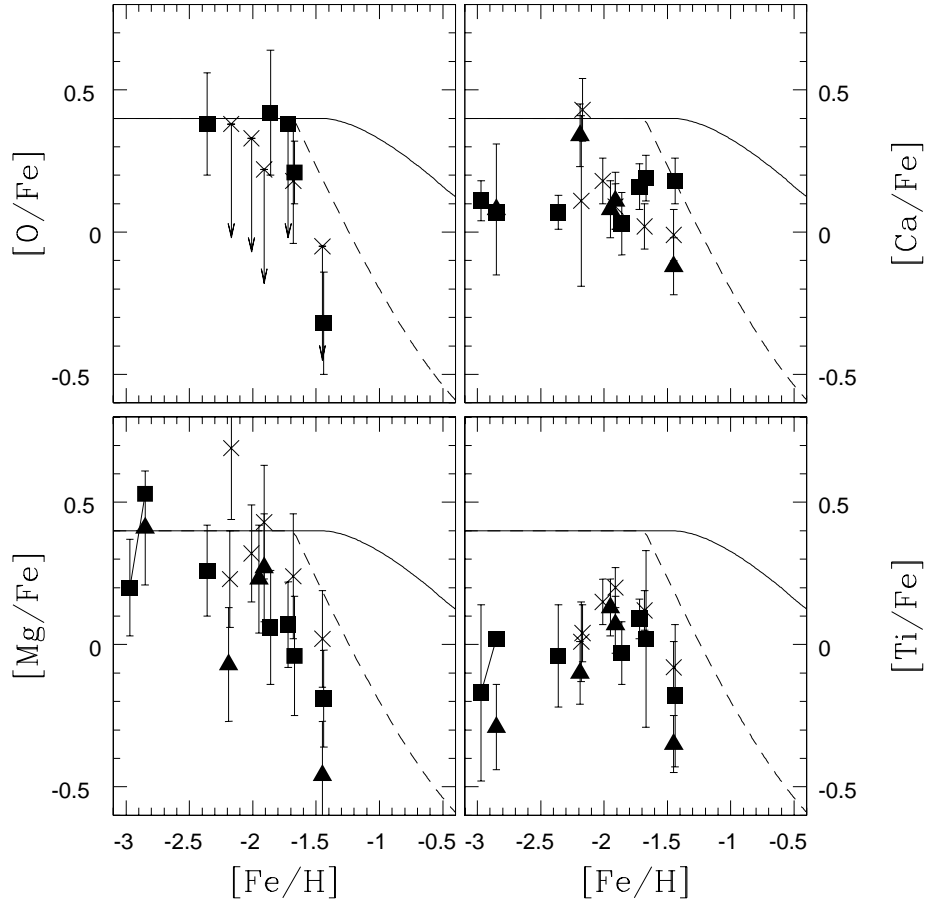

Fig. 1.6. The large symbols with error bars are the same as in Fig. 1.4. The large filled square without an error bar connected to the large filled square with an error bar is D119 from Fulbright et al. (2004). The solid and dashed lines are same as in Fig. 1.1.

A comparison of the Scl abundance patterns with the literature dSph abundances for the other simple dSph galaxies reveals remarkable uniformity. One exception is the neutron capture elements. Unfortunately, the Eu abundances from Shetrone et al. (2001) do not include HFS and thus should be corrected before using them. For example, the most metal-rich star in Ursa Minor should have a HFS correction of 0.26 dex downward, which would bring [Ba/Eu] to −0.46 dex. Once corrected, two of the remaining three simple dSph systems can be fit with a similar toy model with a rapid decline in the star formation rate, but not necessarily at the same metallicity or of the same magnitude. The third simple system, Ursa Minor, exhibits a fairly flat and low [Ba/Eu] abundance trend. While this implies the neutron capture elements are dominated by the *r*-process, we should also note that the most metal-rich Ursa Minor star mentioned above has a very large *r*-process enhancement in comparison to MW halo stars of similar metallicity. We prefer not to draw a conclusion about the Ursa Minor galaxy neutron capture history because of the small sample size and the odd features of this star.

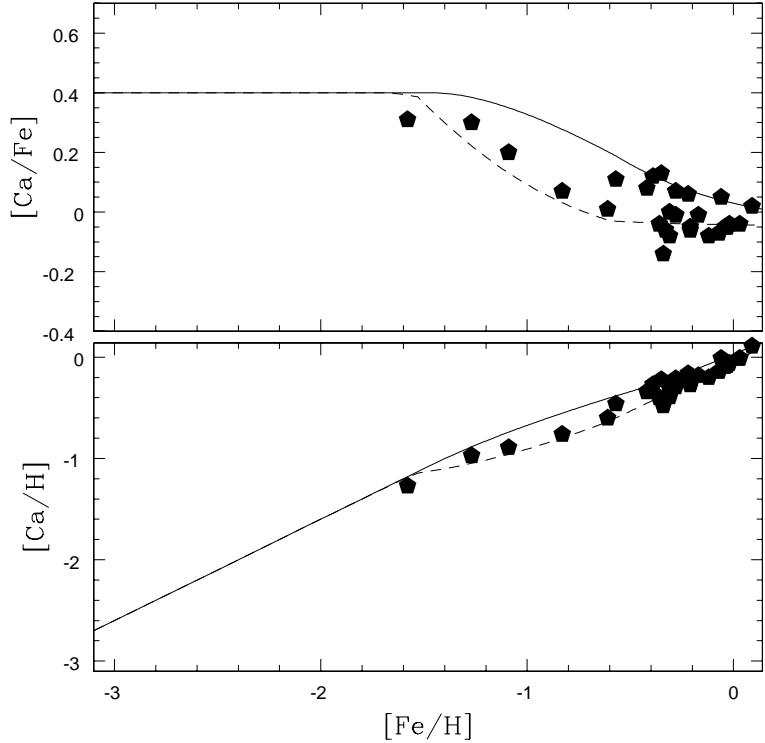

Fig. 1.7. The filled pentagons are from Bonifacio et al. (2000, 2004), Bonifacio & Caffau (2003), and Smecker-Hane & McWilliam (2004). The Ca abundances from Bonifacio have been corrected upward. The solid lines represent the MW toy model and WM mean. The dashed line represents a Sgr toy model.

### 1.4 Complex Systems

The best studied among the complex systems is Sagittarius (Sgr), with two stars from Bonifacio et al. (2000), 14 stars from Smecker-Hane & McWilliam (2004), and 10 stars from Bonifacio & Caffau (2003) and Bonifacio et al. (2004). We begin by looking in detail at this galaxy as a template for the other complex systems. In Figure 1.7 the Ca abundances and abundance ratios are plotted against the metallicity. The solid line represents the MW toy model described previously. On average the Ca abundances fall below the MW toy model at every metallicity. The dash line represents a Sgr toy model constructed in a way similar to the Scl toy model described in the previous section. The Sgr toy model has a initial burst followed by a 25% decline in the star formation rate, and then a second burst of equal strength to the initial burst. These, values were chosen to fit the Ca abundances, and this model is for purely illustrative purposes. The Sgr toy model predictions for the Ca abundance ratios exhibit a decline at [Fe/H] = −1.5, as the SNe Ia begin to contribute extra Fe. The Ca abundance ratios then level out as the second burst adds extra SN II-minted Ca. In contrast to the split seen between light $\alpha$ elements and heavy $\alpha$ elements in the most metal-poor stars among the simple dSph galaxies, the most metal-poor star in Sgr has high Ca and Ti abundances (+0.31 and +0.33, respectively). This seems to imply that Sgr has initial chemical evolution conditions more similar to the MW than to the simple dSph galaxies.

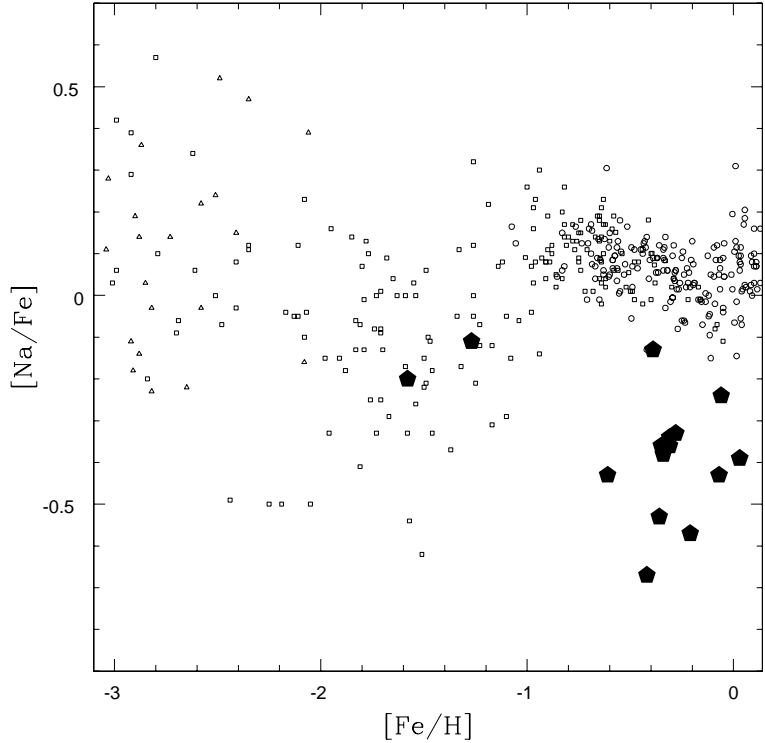

Fig. 1.8. The large filled pentagons are from Bonifacio et al. (2000) and Smecker-Hane & McWilliam (2004). The small symbols are the MW sample: Gratton & Sneden (1988), Edvardsson et al. (1993), McWilliam et al. (1995), Prochaska et al. (2000), and Fulbright (2002).

One metal-poor Sgr star observed by Smecker-Hane & McWilliam (2004) exhibits the deep mixing abundance pattern, significantly enhanced Al and Na (see Kraft 1994 for a review). While this is only a single star, it should be noted that out of the hundreds of MW halo stars and tens of simple dSph galaxy stars surveyed none exhibit the deep mixing abundance pattern. If a significant fraction of the metal-poor stars exhibit the deep mixing pattern then this could place interesting constraints on the formation of the MW halo. Further discussion will have to wait for more extensive surveys of the metal-poor Sgr field and globular cluster stars.

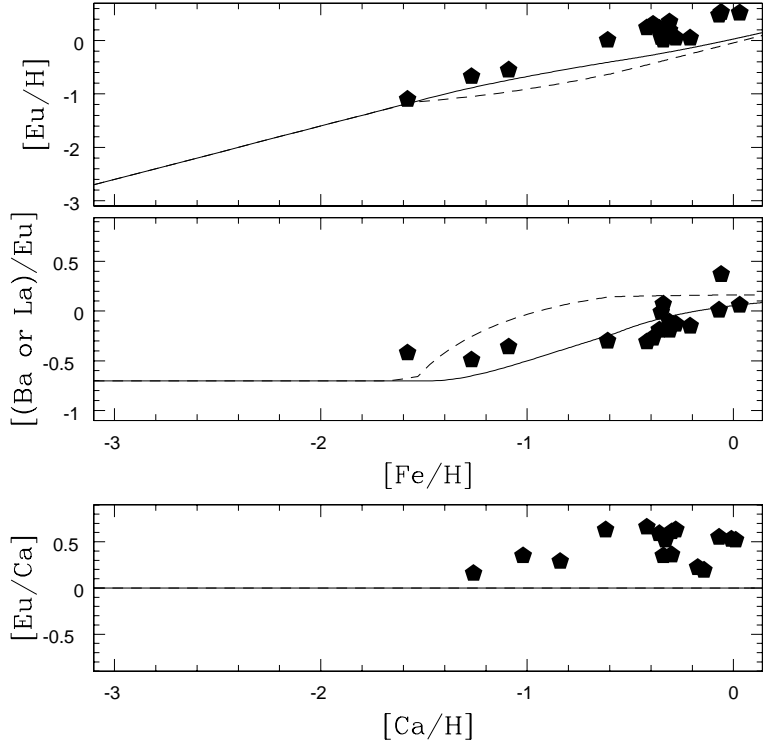

Fig. 1.9. The symbols are the same as in Fig. 1.7.

Figure 1.8 shows the Na abundance ratios for Sgr dSph stars, excluding the star that exhibits the deep mixing pattern mentioned previously. The two remaining metal-poor Sgr stars have abundances similar to those found in the MW halo. This is in stark contrast to simple dSph stars that have systematically lower Na abundances. The metal-rich Sgr stars have Na abundance ratios lower than those found in the MW disk. These low Na abundance ratios are similar to those found in the simple dSph stars despite the lower metallicity of the simple dSph stars. As mentioned earlier, Smecker-Hane & McWilliam (2004) have attributed the low Na abundances in the Sgr dSph galaxy to an excess of Fe from SNe Ia, which should not produce any Na (e.g., Thielemann et al. 1986). The excess of Fe produced from SNe Ia is consistent with the low $\alpha$ abundances in these metal-rich dSph stars.

The predictions for the neutron capture elements are shown in Figure 1.9. The Sgr toy model was created using the same prescription for chemical evolution as used in both the

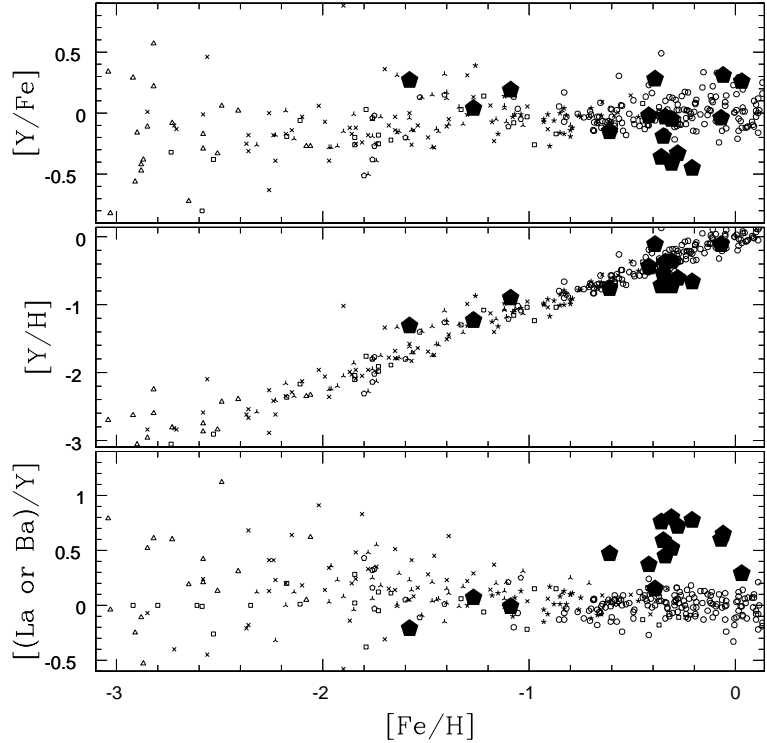

Fig. 1.10. The large pentagons are the same as in Fig. 1.7. The small symbols are the MW sample and the same as in Fig. 1.4.

Scl and MW toy models. The Sgr toy model underpredicts the amount of Eu actually found in the metal-rich Sgr stars. At metallicity [Fe/H] = −0.5 the toy model predicts too large a contribution from AGB star *s*-processed material than is actually detected in the stars. This may only indicate that the metallicity at which the AGB stars begin to contribute La is higher than modeled—that the early chemical evolution was faster than modeled and more similar to the MW. Other interpretations of this poor fit of the model could be due to poor modeling of the *s*-process (e.g., a lack of metal-dependent yields), or the anomalously large Eu abundances. At the solar ratio of *s*-process to *r*-process material ([La/Eu] = 0 ), Eu has a tiny ∼ 5% contribution from the *s*-process. This amounts to a meager 0.02 dex increase to a pure *r*-process Eu abundance. Thus, the overabundance of Eu is not likely to be caused by an *s*-process contribution for the metal-rich Sgr stars, since their [La/Eu] ratios are approximately 0. To investigate this further, we plot the Eu to Ca abundance ratio against Ca abundance in the bottom pane of Figure 1.9. As mentioned previously this ratio is roughly zero at all metallicities in the MW, while in the Sgr dSph it is super solar at all metallicities with a small trend toward larger ratios at higher metallicities. In our toy model the Eu abundances are made in the same proportion with the $\alpha$ elements. A divergence between the $\alpha$ elements and the *r*-process elements seen in this and perhaps other dSph may be a clue to the origins of the *r*-process site.

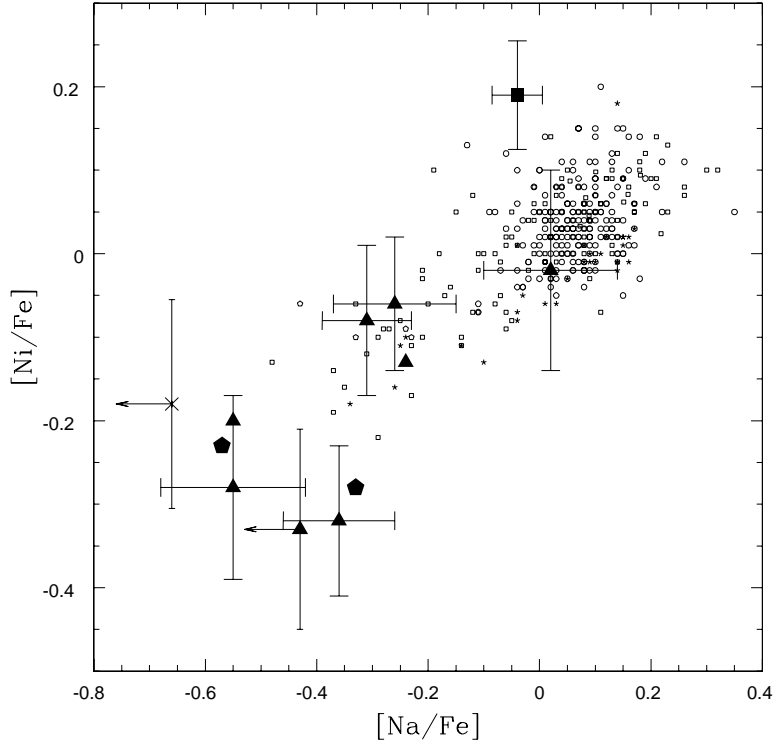

Fig. 1.11. The large symbols the dSph sample with the metallicities in the range $-1.5 <$ [Fe/H] $< 0.0$ from: Bonifacio et al. (2000), Shetrone et al. (2001, 2003), and Geisler et al. (2004). The small symbols are from the MW sample in the same metallicity range: Gratton & Sneden (1988), Edvardsson et al. (1993), Nissen & Schuster (1997), Stephens (1999), Prochaska et al. (2000), Fulbright (2002), and Stephens & Boesgaard (2002).

Figure 1.10 exhibits the Sgr Y abundances, abundance ratios, and the heavy *s*-process to light *s*-process ratio ([La/Y]). The metal-poor Sgr stars have [La/Y] ratios that are similar to those of the halo; however, at this metallicity the La is mostly of *r*-process origin. The [La/Y] ratio among the more metal-rich Sgr stars is significantly enhanced with respect to the MW stars of similar metallicity. Smecker-Hane & McWilliam (2004) argue that the underabundance of Y is due to the metallicity-dependent origin of Y and La *and* a delay in the incorporation of the AGB *s*-process yields. They argue that the metal-rich generation of AGB stars did not contribute to the chemical evolution of that generation, but rather, the high [La/Y] ratio suggests the contribution from low-metallicity AGB stars. If true, this breaks the instant-recycling rule used in many chemical evolution models. Among the metal-rich Sgr stars [Y/Fe] is only slightly lower than the MW mean at the same metallicity. However, the lower panel of Figure 1.9 suggests that there is a high *r*-process contribution at a given metallicity and La has both an *r*-process and *s*-process contribution, so some of the enhanced [La/Y] contribution could be due to an enhanced *r*-process contribution and unrelated to the *s*-process.

The sample sizes for the other complex dSph stars are very small, but comparing the trends seen in the dSph with those found in Leo I, Carina, and Fornax reveals some significant differences: (1) The metal-poor stars exhibit Na abundance ratios similar to the simple dSph stars and below those found in the Sgr dSph; (2) the single metal-rich star in Fornax has enhanced $\alpha$ abundance ratios with respect to the halo and the Sgr dSph.

## 1.5 Connection to the Milky Way

Several lines of evidence suggest that the Galactic halo is, at least partially, composed of accreted dwarf galaxies. These include the current assimilation of the Sgr dwarf (Ibata et al. 1997; Dohm-Palmer et al. 2000; Newberg et al. 2002), and possibly $\omega$ Cen (Majewski et al. 2000, assuming that it is a stripped dwarf galaxy). Since the halo's metallicity distribution peaks near [Fe/H] = $-1.8$ and those stars show a higher heavy $\alpha$ to iron ratio than the dSph stars, clearly a large percentage of the halo cannot have been produced from dSphs similar to the simple systems reviewed in this work. In Fulbright (2002), fewer than 10% of the local metal-poor ([Fe/H] $< -1.2$) stars in that sample have heavy $\alpha$ to iron abundance ratios similar to those found in the dSph sample. However, by subdividing that sample by total space velocity, Fulbright found that the highest-space velocity stars have systematically lower $\alpha$ to iron abundance ratios. Nissen & Schuster (1997) conducted a detailed abundance analysis of a nearby sample of disk and halo stars with similar metallicities to study the disk-halo transition. Their sample was chosen to get an equal number of disk and halo stars as defined by the stars' rotation. Of their 13 chosen halo stars, eight show an unusual abundance pattern: low $\alpha$ element to iron ratio, low [Ni/Fe] abundances, and low [Na/Fe] abundances. These odd halo stars also exhibited larger $R_{max}$ and $z_{max}$ orbital parameters than the other halo stars sampled. Nissen & Schuster (1997) suggest that these anomalous stars may have their origins in disrupted dSph. In Figure 1.11 the dSph stars with metallicities between $-1.5$ and $-0.5$ are shown along with literature MW field stars in the same metallicity range. The dSph stars systematically fall among the stars that Nissen & Schuster (1997) predicted may have an origin in dSph galaxies. This seems to lend support to the idea put forward by Nissen & Schuster (1997) that a large fraction ($> 50$%) of the metal-rich halo stars may have their origin in disrupted dSph like those studied in this work. This still leaves the question of the origin of the metal-poor halo stars and the fraction of the metal-poor halo stars that formed through monolithic collapse versus accretion of dSph galaxies.

$\omega$ Cen is a system once categorized as a globular cluster, but recent studies suggest that it may actually be a captured galaxy (e.g., Majewski et al. 2000). $\omega$ Cen exhibits a large metallicity spread, with the more metal-rich stars exhibiting very large *s*-process element enhancements, implying an age spread and self-pollution. Unlike the simple dSph systems, a large fraction of the metal-poor $\omega$ Cen giants exhibit a deep mixing abundance pattern. Could the Sgr dSph be similar to $\omega$ Cen in this regard?

## 1.6 A History of Conclusions

(1) Most dSph abundance patterns do not look like that of the average MW halo; thus, it would be difficult to make the majority of the MW halo from building blocks that look like today's dSph galaxies (Shetrone et al. 1998, 2001).

(2) Abundances derived from color-magnitude diagrams can be dramatically fooled by the age-metallicity relationship, as seen in the Sgr dSph. The Sgr dSph may be much younger and more metal rich that previously thought (Bonifacio et al. 2000).

(3) Studies of the outer halo suggest that it may have abundances similar to those found in the simple dSph galaxies (Nissen & Shuster 1997; Fulbright 2002; Stephens & Boesgaard 2002).

(4) Both complex and simple dSph systems show significant evolution from a SN II abundance pattern to a SN Ia abundance pattern (Shetrone et al. 2001, 2003; Smecker-Hane & McWilliam 2004).

(5) Evidence from metal-poor dSph stars with a clear indication of SN Ia contamination indicates that Mn and Cu are not produced in large quantities in metal-poor SNe Ia (McWilliam et al. 2003; Shetrone et al. 2003).

(6) There appears to be a split between the light and heavy $\alpha$ elements in the most metal-poor simple dSph stars (Fulbright et al. 2004; this work).

(7) The Sgr dSph metal-poor stars do not exhibit the split between the light and heavy $\alpha$ elements, suggesting abundances more like those found in the MW globular clusters, including a star exhibiting a deep mixing abundance pattern (Smecker-Hane & McWilliam 2004).